# Understanding Search and Decision Errors in Liver Metastasis Detection and the Effects of Lower Radiation Dose

Parmita Mondal,[a] Rickey Carter,[b] Andrew D Missert,[a] Akitoshi Inoue,[a] Joel G. Fletcher,[a] Hao Gong,[a] Shuai Leng,[a] Lifeng Yu,[a] Cynthia H. McCollough,[a] and Scott S. Hsieh[a]

[a]Mayo Clinic, Department of Radiology, Rochester, Minnesota, United States
[b]Mayo Clinic, Department of Quantitative Health Sciences Research, Jacksonville, Florida, United States

**Background:** The detection performance of liver metastases decreases with the reduction of radiation dose, but misses are heterogeneous. Previous eye tracking work has characterized missed metastases into two categories: search errors i.e., the eyes never land on the lesion, and decision errors i.e., the lesion is seen but not recognized as malignant.

**Purpose:** To measure how frequently radiologists miss liver metastases at reduced radiation dose, when metastases are categorized as those that are prone to search errors ("search-error-dominant") or those that are prone to decision errors ("decision-error-dominant").

**Methods:** We integrated three prior reader studies to answer this question. In all studies, radiologists interpreted the same set of 40 contrast-enhanced abdominal CT exams containing 91 liver metastases whose locations had been previously marked. In two studies, the workstation recorded their gaze and eye movements. Using eye dwell times, metastases were classified as search-error-dominant (majority of misses had <2 sec gaze time) or decision-error-dominant (>2 sec gaze time). In the third study, exams were interpreted both at 120 and 200 quality reference mAs (QRM) by ten radiologists. The third study did not include eye tracking. Out of 91 liver metastases, we excluded 16 that were never missed in the eye tracking studies and used 75 liver metastases for the present study.

**Results:** Data from the first and second studies and a gaze threshold of 2 seconds were used to separate search from decision errors. 49.3% (37/75) metastases were categorized as search-error-dominant and 50.7% (38/75) as decision-error-dominant metastases. In the third study, 20.8% (156/750) of metastases were missed at 120 QRM and 14.8% (111/750) of metastases were missed at 200 QRM. Decision-error-dominant metastases were missed at similar frequency at 120 QRM as in 200 QRM (10% vs. 7.63% respectively, $p = 0.53$), but search-error-dominant metastases were more frequently missed at lower dose (31.62% vs. 22.16%, $p = 0.002$). Differences were weaker when a gaze threshold of 1 second was used.

**Conclusion:** By integrating a large eye tracking dataset, we characterized lesion misses by separating them into search-error-dominant and decision-error-dominant metastases. This framework allowed us to evaluate how radiation dose affect lesion detection across these error categories. The findings were sensitive to the definition of gaze threshold that was used.

## Introduction

A central task for the radiologist is lesion detection (1). Lesion detection is impacted by several factors, including lesion contrast and conspicuity, the radiation dose of the acquisition protocol, the use of advanced reconstruction or post-processing to reduce noise, and the performance of the individual radiologist. It is known that when radiation dose is reduced, image noise increases, and this obscures subtle metastases such as low-contrast metastases. For example, Goenka et al. showed in a phantom model with 36 model metastases and 18 readers that detection performance was noninferior at 25% dose reduction, but inferior at 50% dose reduction even with iterative reconstruction (2). In studies like these, metastases are unambiguous and the only task is visual search. In clinical practice, metastases must further be classified into benign or malignant entities, and this adds another point of failure (3, 4).

Pooler et al. performed a study that included 70 patients with nonprimary hepatic malignancies who underwent routine-dose and lower-dose (60-70% dose reduction) abdominal CT with filtered back projection (FBP) (5). They found that using iterative reconstruction (IR) at the lower dose did not preserve observer performance. Mileto et al. reviewed that IR helps reduce noise in abdominal CT. However, for low-contrast tasks using real patient and phantom data, lesion detectability is generally preserved only with dose reduction of about 25%, while larger reductions can degrade performance (6). This was further confirmed by Jensen et al., who performed a prospective study that included 52 adults with 233 colorectal liver metastases who underwent routine-dose and reduced-dose scans where observer performance at routine-dose FBP and reduced-dose was compared. Lower dose CT examinations using IR demonstrated decreased observer performance (7). Fletcher et al. examined contrast-enhanced abdominal CT in the detection of hepatic metastases and found that at moderate levels of dose reduction (40%), observer performance was the same regardless of whether IR or FBP was employed, while reader confidence declined for both FBP and IR for small low-contrast metastases (8, 9). Therefore, in previous studies, it is well established that lower dose reduces performance. However, previous studies do not differentiate why metastases are missed.

Our study addresses this gap by using eye tracking data to separate missed metastases into two categories: those missed predominantly by search errors and those missed predominantly by decision errors. A search error occurs when the radiologist’s eyes never actually land on the metastases, or when the eye gaze is so brief that there is no conscious recognition of a possible lesion. In contrast, a decision error occurs when the lesion is fixated upon for a period of time, but the radiologist chooses to categorize it as benign when it is malignant, reflecting a cognitive or interpretive failure rather than a perceptual one (10, 11).

Hsieh et al. (12) examined reader variability at routine dose only while using eye-tracking hardware and found that abdominal subspecialists demonstrated higher diagnostic discrimination (area under the jackknife alternative free-response receiver operating characteristic curve, JAFROC AUC) = 0.77) than trainees (JAFROC AUC = 0.71) or non-abdominal radiologists (JAFROC AUC = 0.69), though sensitivity was similar across groups. They also found that higher sensitivity correlated with longer interpretation time and greater use of coronal images. Routine dose in this cohort was 200 quality reference mAs (QRM). In a follow-up investigation, Hsieh et al. (13) tested the effect of a training program on 31 readers using the same 40 CT cases, showing that search errors decreased significantly (10.8% to 8.1%, $p = 0.01$), but classification errors and overall diagnostic accuracy did not improve. Gains were most pronounced among trainees and non-abdominal radiologists, suggesting that targeted training can reduce oversight errors, whereas classification skill likely requires longer or more intensive experience. Fletcher et al. examined reader performance at different dose levels with a cohort of ten abdominal radiologists and found that performance diminished at lower dose (8, 9).

Eye tracking measures precisely where and for how long a reader's gaze lingers on specific image areas and thus provides a way to classify missed metastases into these two categories. Eye tracking requires the experimenter to select a time threshold to classify missed metastases into either search or decision errors (14). In this study, missed metastases in which a reader's gaze fixates near a lesion for less than 2 sec are coded as a search error, while longer fixations are coded as a decision error (15, 16).

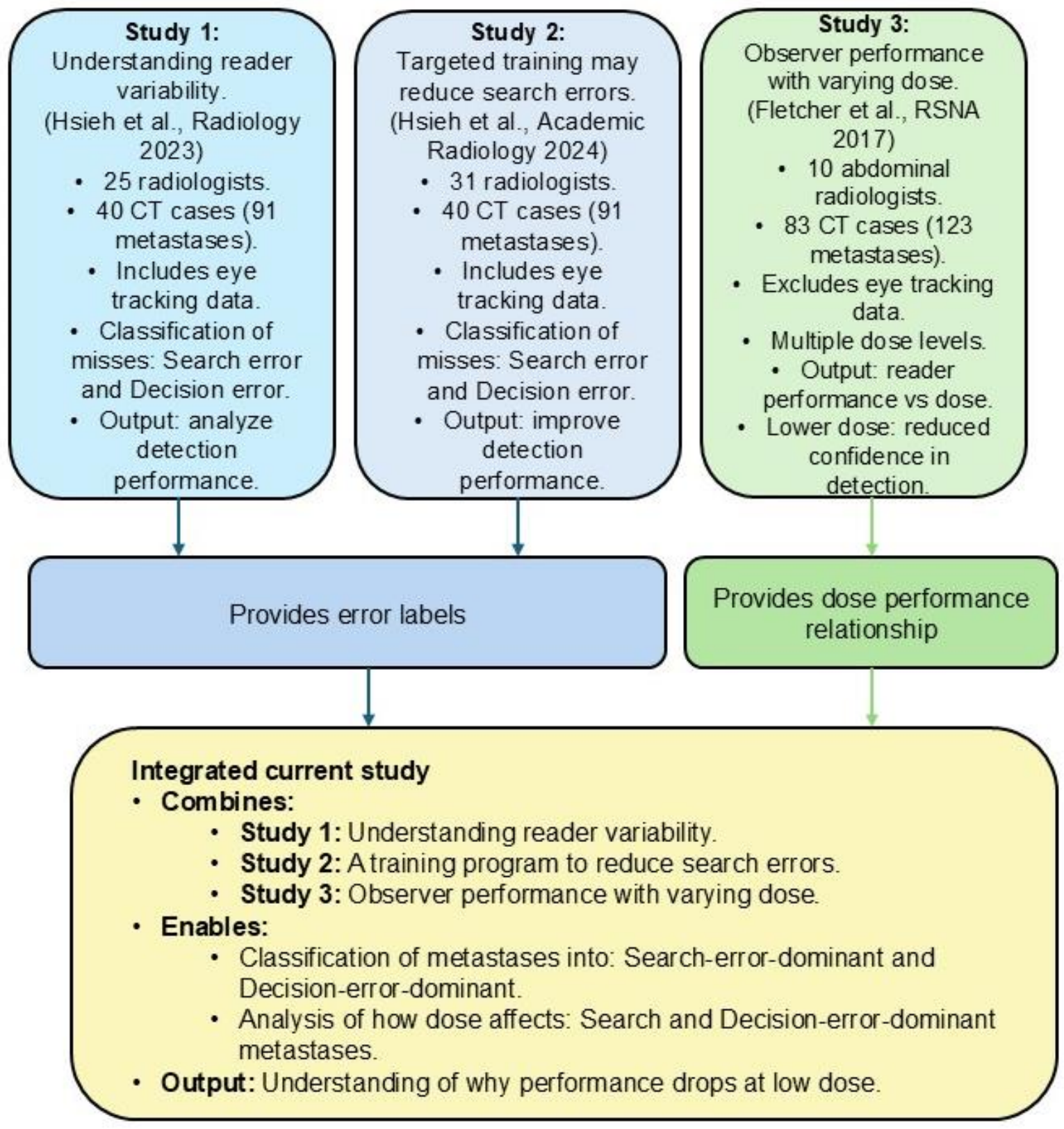

**Figure 1.** shows the flowchart of the three studies, integrated into the present study.

## Material and Methods

Acquisition of the data used in this study was approved by our institutional review board (IRB). This work is a reanalysis of three separate studies in which readers examined the same set of 40 cases (32 containing 91 liver metastases and eight normal).

The first two studies (12, 13) included eye tracking and were used only to classify liver metastases into search-error-dominant and decision-error-dominant. The third study did not include eye tracking but included multiple dose levels, and this was used to evaluate reader performance at lower dose (8). The third study also included other dose levels and cases that are not analyzed here; we analyzed only 200 QRM and 120 QRM reconstructed with FBP, which represent routine dose and a clinically relevant reduced dose, respectively. This allowed us to test whether lower dose affects search-error-dominant and decision-error-dominant metastases while keeping the comparison simple and interpretable. These three prior studies are integrated together into this present study, which is a reanalysis of their data. Out of the 91 liver metastases, 16 were not missed in either of the two eye tracking studies and were excluded, leaving 75 liver metastases available for this present study.

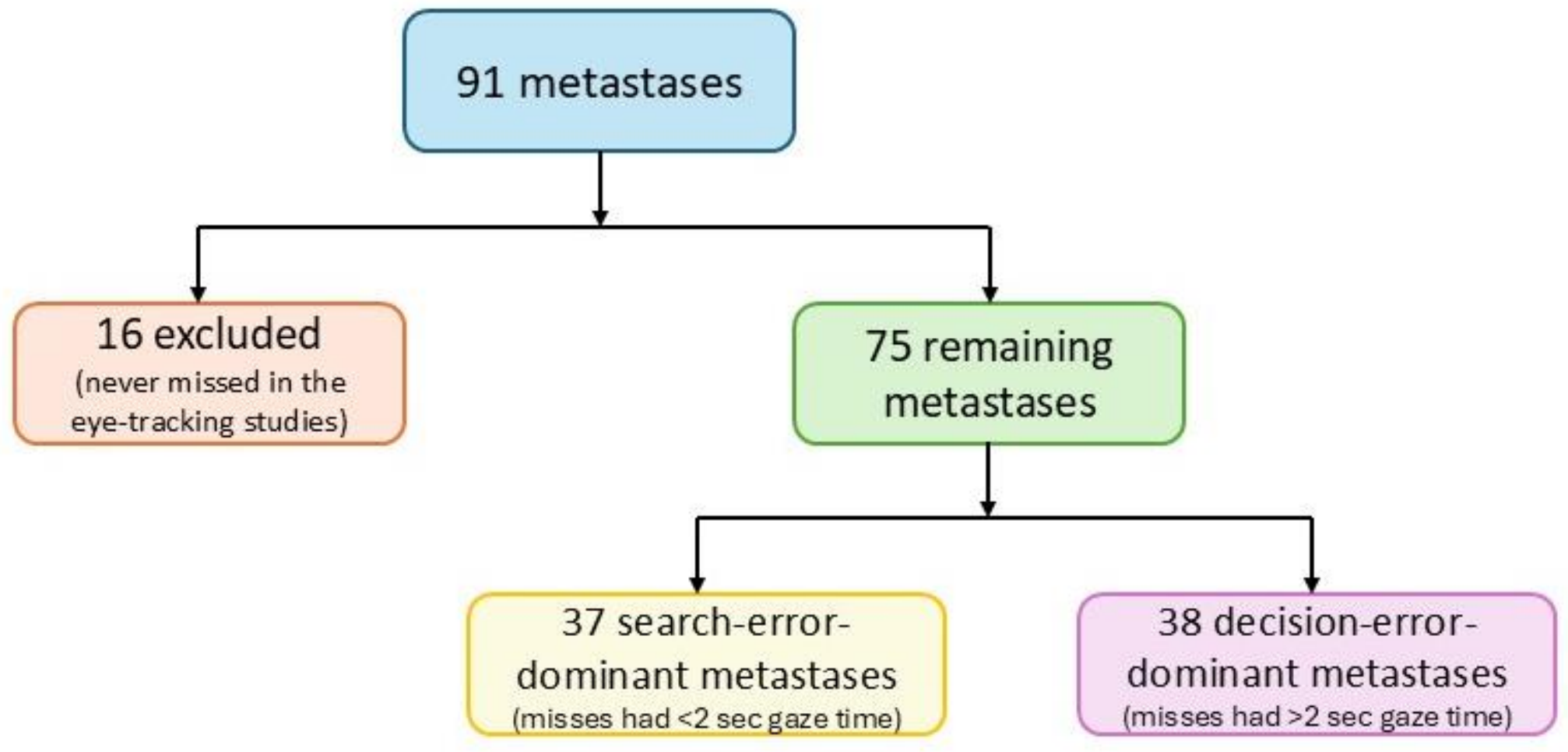

**Figure 2.** shows the flowchart of metastasis selection and error type classification: search-error-dominant (majority of misses had <2 sec gaze time) and decision-error-dominant (>2 sec gaze time) for the primary analysis of 2sec threshold.

**Imaging Protocol:**

The imaging protocol has been described in previous work (17-19). Briefly, iodine-enhanced abdominal CT scans were performed using 64- or 128-section Siemens scanners (Definition, Definition Flash, or Definition AS+; Siemens Healthcare) at routine radiation dose levels of approximately 200 quality reference milliampere-seconds (QRM). In the third (multidose) study, projection-level noise insertion was used to synthesize 120 QRM reconstructions. Abdominal CT images from the 40 patients were obtained in the portal venous phase and reconstructed with 3-mm slice thickness at 2-mm intervals using FBP and a medium-smooth (B30f) kernel. Ground truth for liver metastases was confirmed by histopathological analysis or radiologic evidence of progression. Readers were asked to mark all suspected metastases. A metastasis was classified as detected when its marking overlapped with the ground-truth annotation; otherwise, it was missed.

**Lesion Classification:**

We used a 500 Hz eye tracker (Eyelink Portable Duo; SR-Research) with typical accuracy equivalent to about 15 mm in anatomy to record gaze positions mapped onto Digital Imaging and Communications in Medicine (DICOM) coordinates (12). Gazes near a lesion within 40 pixels and two CT slices were summed (20). This allowed quantification of how much attention each lesion received and whether visual engagement translated into detection.

Missed metastases were categorized according to this total gaze time. If the total gaze time exceeded 2 seconds, the reader was coded to have made a decision error; otherwise, it was coded to be a search error. Metastases were labeled as search-error-dominant if the number of search errors was more than the number of decision errors and decision-error-dominant if the number of decision error was greater than or equal to the number of search errors. While most of our analyses use a 2-second gaze threshold, we tested a 1-second gaze threshold in a sensitivity analysis. Figure 1 shows the flowchart of the three studies, integrated into the present study and Figure 2 shows the flowchart of metastasis selection and error type classification.

Misses were categorized by a total of N = 56 readers, including 25 readers from the first study (12) and 31 readers from the second study (13). The second study included a pre-training read and a post-training read, but only the pre-training read was used in this analysis because of possible memory effects between the pre-training read and the post-training read (washout period of 2-5 weeks). Some readers participated in both studies, but the time interval between these studies was about one year, making memory effects unlikely.

**Statistical Analysis:**

We used generalized estimating equations (GEE) to model the probability of lesion detection while accounting for the clustered structure of the data, since multiple observations were contributed by the same metastases and radiologists. GEE is appropriate for correlated binary outcomes and allows estimation of the effects of lesion category, radiation dose, and their interaction on detection performance (21). Clustering was performed at the metastasis level, not at the radiologist level. This approach allowed us to quantify differences in detection and miss rates across conditions while properly accounting for within-cluster correlation in the reader study data. In addition, we also computed estimated marginal means (emmeans) from the fitted GEE model to obtain predicted detection probabilities and to facilitate comparisons between lesion categories and dose levels. All analyses were performed using R version 4.5.2 (R Foundation for Statistical Computing, Vienna, Austria).

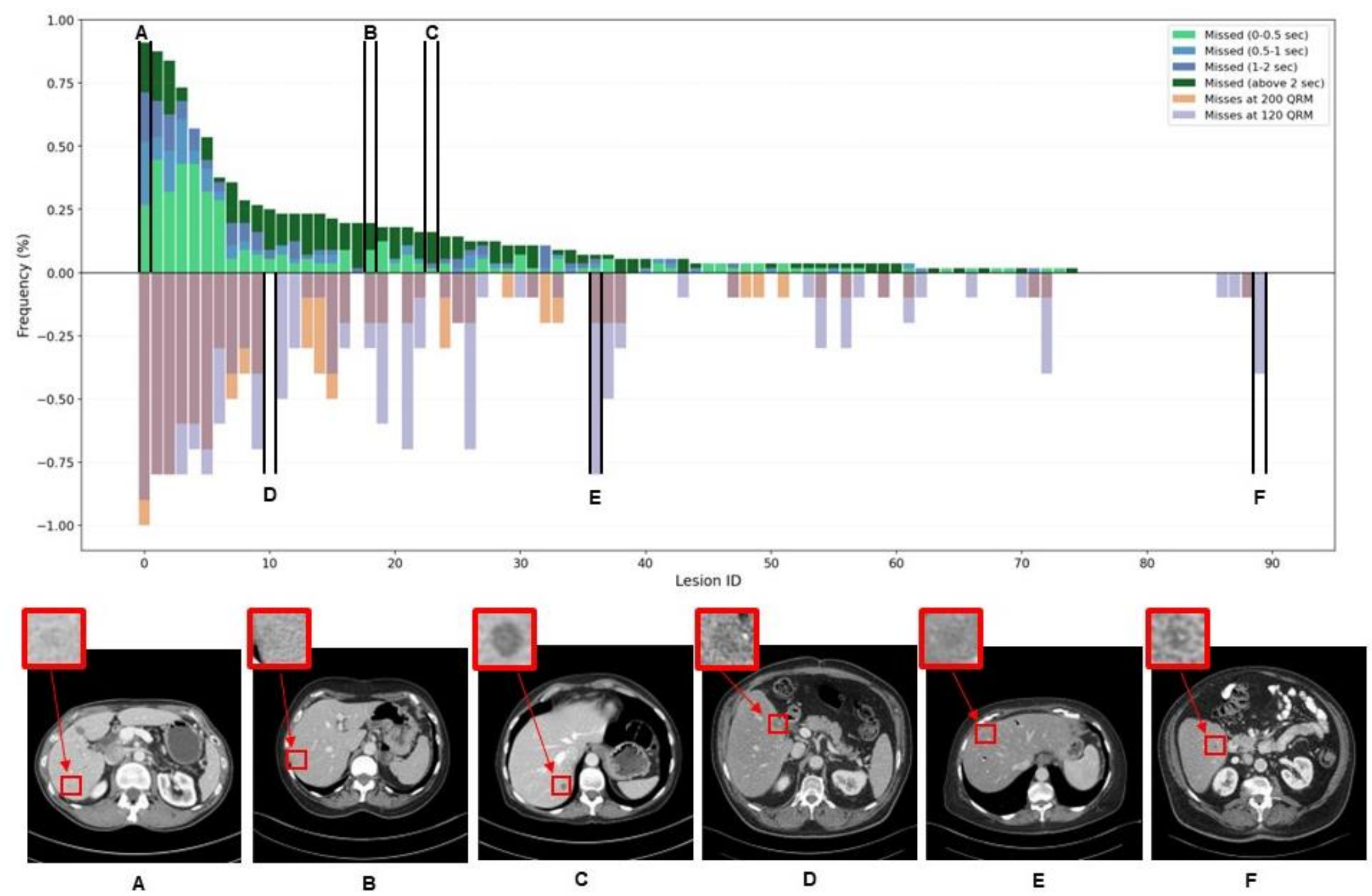

**Figure 3.** Bar plot showing distribution of data, where x axis is each lesion and y axis gives the percentage of misses by radiologist for each lesion. The top part of the plot shows the misses at 0-0.5 seconds, 0.5-1 seconds, 1-2 seconds and above 2 seconds. The bottom part shows the misses at 200 and 120 QRM. The data in the bottom part was analyzed by 10 subspecialists, while the data in the top part of the plot included 56 readers. Examples below show: (A) search-error-dominant lesion (B) equal search and decision error lesion with dose dependency (C) decision-error-dominant lesion (D) decision-error-dominant lesion not missed at 120/200 QRM (E) dose sensitive lesion (F) low miss lesion with dose dependency. Note that the last 15 metastases, including F, were not analyzed in this study because they were not missed in the eye tracking studies.

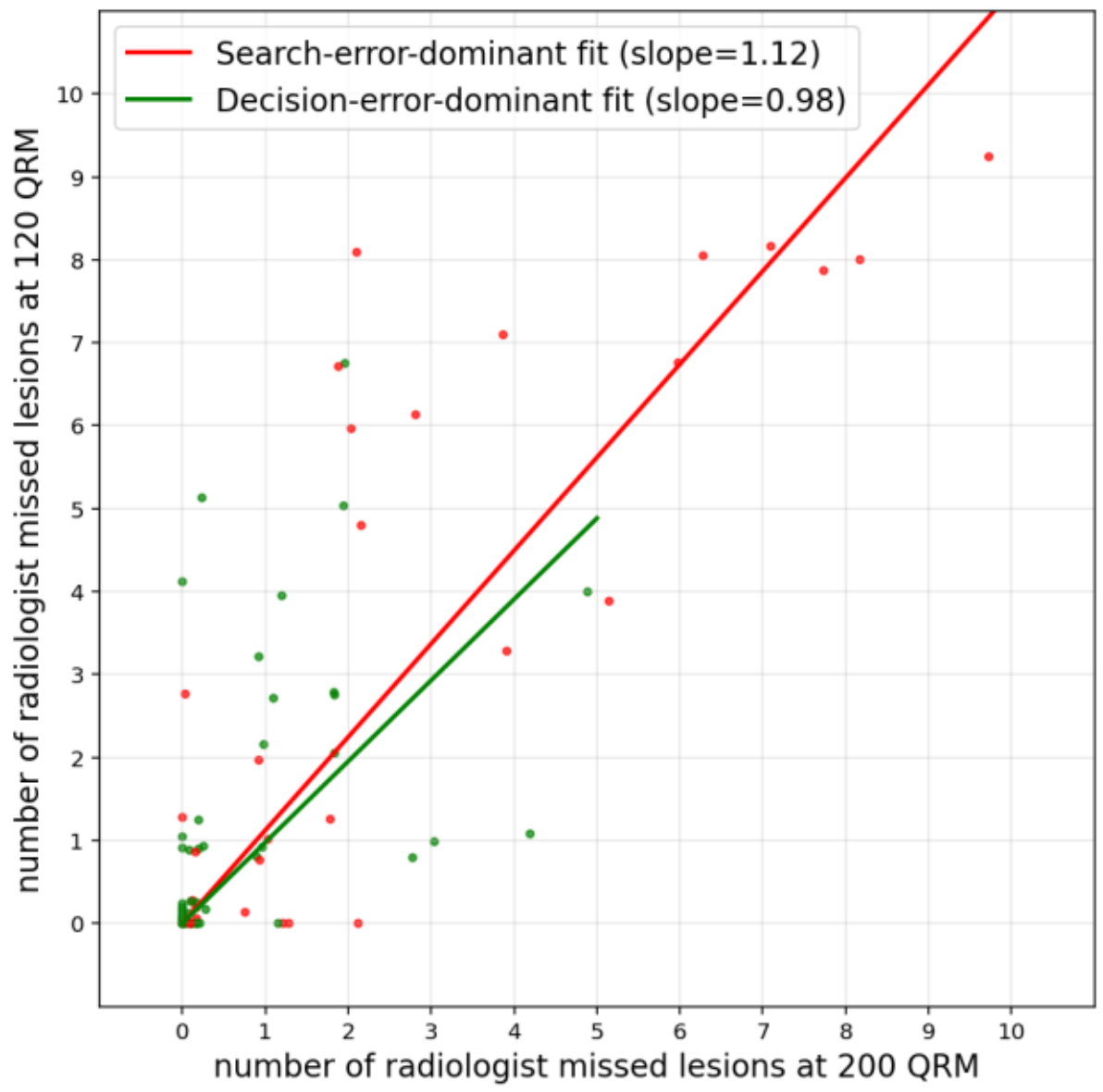

**Figure 4.** Scatter plot showing radiologist miss patterns for gaze threshold 2 seconds across dose levels and error type: search-error-dominant and decision-error-dominant metastases. Points have been jittered to improve readability.

## Results

Figure 3 summarizes our datasets as a combined bar plot, where x axis is each lesion and y axis shows the percentage of misses by radiologist for each lesion. The top part of the plot shows the combined eye tracking studies, from N=56 readers, and categorizes misses according to multiple time thresholds. Our primary analysis uses a 2 second gaze threshold to differentiate search from decision errors, but other thresholds could be used. The bottom part shows the misses at 200 and 120 QRM. Thess data were analyzed by 10 abdominal subspecialist readers. We selected some of the representative metastases from the plot to highlight key insights. Some of the miss rate differences between the top and bottom part of this plot could be explained by differences in the study cohorts: the 120/200 QRM data was read by subspecialists only, while the top part of the plot included trainees also.

When a gaze threshold of 2 seconds was used to separate search from decision errors, 49.33% (37/75) metastases were categorized as search-error-dominant and 50.66% (38/75) as decision-error-dominant metastases. In the multidose study, 20.8% (156/750) of metastases were missed at 120 QRM and 14.8% (111/750) of metastases were missed at 200 QRM. Figure 4 shows that search-error-dominant metastases are missed more frequently at 120 QRM than 200 QRM, while decision-error-dominant metastases are missed with similar frequency. If dose had no effect, the slope of the line would be 1. The slope for decision-error-dominant metastases is 0.98, implying minimal effect from dose, while the slope for search-error-dominant metastases is 1.12, suggesting that reduced dose leads to more frequent misses. However, there is significant heterogeneity among the metastases: many metastases are never missed at either QRM, and a few metastases are missed more than half the time. The intercept of the trend line was fixed at 0 and points have been jittered to help visualize density.

Table 1 shows the distribution of detected and missed hepatic metastases by radiation dose (200 QRM vs 120 QRM) and lesion class (search-error-dominant metastases and decision-error-dominant metastases).

While the slope of the scatterplot is suggestive, a more direct way to test the hypothesis that search-error-dominant (but not decision-error-dominant) metastases are missed more frequently at lower radiation dose is using the GEE analysis. The GEE logistic regression model demonstrated a significant effect of radiation dose on lesion detection. Table 2 shows that higher radiation dose was associated with significantly increased odds of detecting hepatic metastases (odds ratio (OR) = 1.62, 95% CI: 1.19 - 2.21, $p = 0.002$). Figure 5 shows the detection rate across dose levels and error categories: search-error-dominant and decision-error-dominant metastases, respectively.

Our findings are affected by choice of gaze threshold. We used a gaze threshold of 2 seconds to classify missed metastases into search errors or decision errors, but these do not perfectly correspond to the thought processes of the reader: a reader might consciously dismiss a metastasis as a cyst in under 2 seconds, for example, and Figure 3(c) shows that a few readers did this. To avoid limiting our analysis to a single gaze threshold of 2 seconds, we therefore also repeated the categorization using a 1-second gaze threshold. With the new gaze threshold, some search errors were reclassified into decision errors, and 7 metastases shifted from search-error-dominant to decision-error-dominant. Table 3 shows the results with the 1-second gaze threshold obtained from GEE regression, and Figure 6 shows the detection rate for the gaze threshold of 1 second across dose levels and error categories: search-error-dominant and decision-error-dominant metastases, respectively. Table 4 shows the chance of detection at 200 QRM vs 120 QRM. We observe that irrespective of the gaze threshold, the chance of detection at 200 QRM vs 120 QRM for search-error-dominant metastases is significantly different, while, for decision-error-dominant metastases, it is not.

For search-error-dominant metastases at gaze threshold of 2 seconds or 1 second, the odds ratio of detection at 200 QRM versus 120 QRM was 1.62 (95% CI: 1.19, 2.21) or 1.60 (95% CI: 1.17, 2.19), respectively. For decision-error-dominant metastases, the odds ratio of detection at 200 QRM versus 120 QRM was 1.34 (95% CI: 0.80, 2.24) or 1.44 (95% CI: 0.94, 2.21), respectively. While the odds ratio for decision-error-dominant metastases were not significantly different from 1, they do still show a trend towards increased detection at higher dose that might become significant with a larger sample size.

**Table 1.** Distribution of detected and missed hepatic metastases for a gaze threshold of 2 seconds by radiation dose (200 QRM vs 120 QRM) and lesion class (search-error-dominant metastases and decision-error-dominant metastases), where N is total number of observations in that group. QRM= quality reference milliampere-seconds

Search-error-dominant metastases

| Dose (QRM) | Number of metastases | Percentage of misses (%) | Total markings |
|---|---|---|---|
| **120** | 37 | 31.62 (117/370) | 370 |
| **200** | 37 | 22.16 (82/370) | 370 |

Decision-error-dominant metastases

| Dose (QRM) | Number of metastases | Percentage of misses (%) | Total markings |
|---|---|---|---|
| **120** | 38 | 10 (38/380) | 380 |
| **200** | 38 | 7.63 (29/380) | 380 |

**Table 2.** Results for gaze threshold 2 seconds obtained from generalized estimating equations (GEE), where CI is confidence interval and class represented search-error or decision-error dominant misses

| Variable | Estimate | Odds ratio (CI lower, CI upper) |
|---|---|---|
| **Intercept** | 0.77 | 2.16 (1.33, 3.50) |
| **Class** | 1.42 | 4.16 (2.11, 8.21) |
| **Dose** | 0.48 | 1.62 (1.19, 2.21) |
| **Interaction** | -0.18 | 0.82 (0.45, 1.50) |

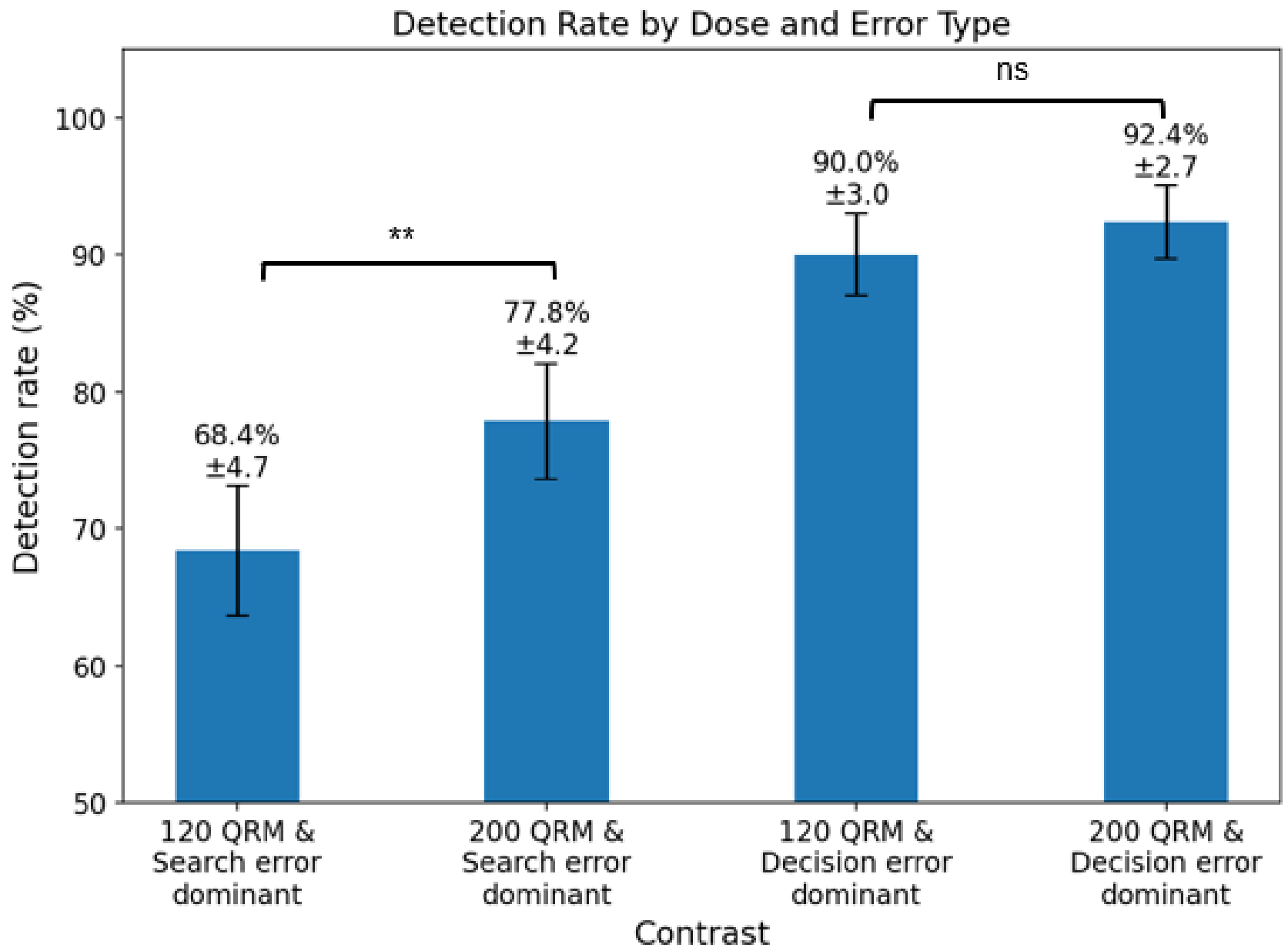

**Figure 5.** shows detection rate for gaze threshold 2 seconds across dose levels and error categories: search-error-dominant and decision-error-dominant metastases, respectively, where ** represents p-value < 0.01 and ns represents not significant. QRM= quality reference milliampere-seconds.

**Table 3.** Results for gaze threshold 1 second obtained from generalized estimating equations (GEE), where CI is confidence interval.

| Variable | Estimate | Odds ratio (CI lower, CI upper) |
|---|---|---|
| **Intercept** | 0.83 | 2.29 (1.30, 4.03) |
| **Class** | 0.94 | 2.57 (1.24, 5.35) |
| **Dose** | 0.47 | 1.60 (1.17, 2.19) |
| **Interaction** | -0.10 | 0.90 (0.53, 1.53) |

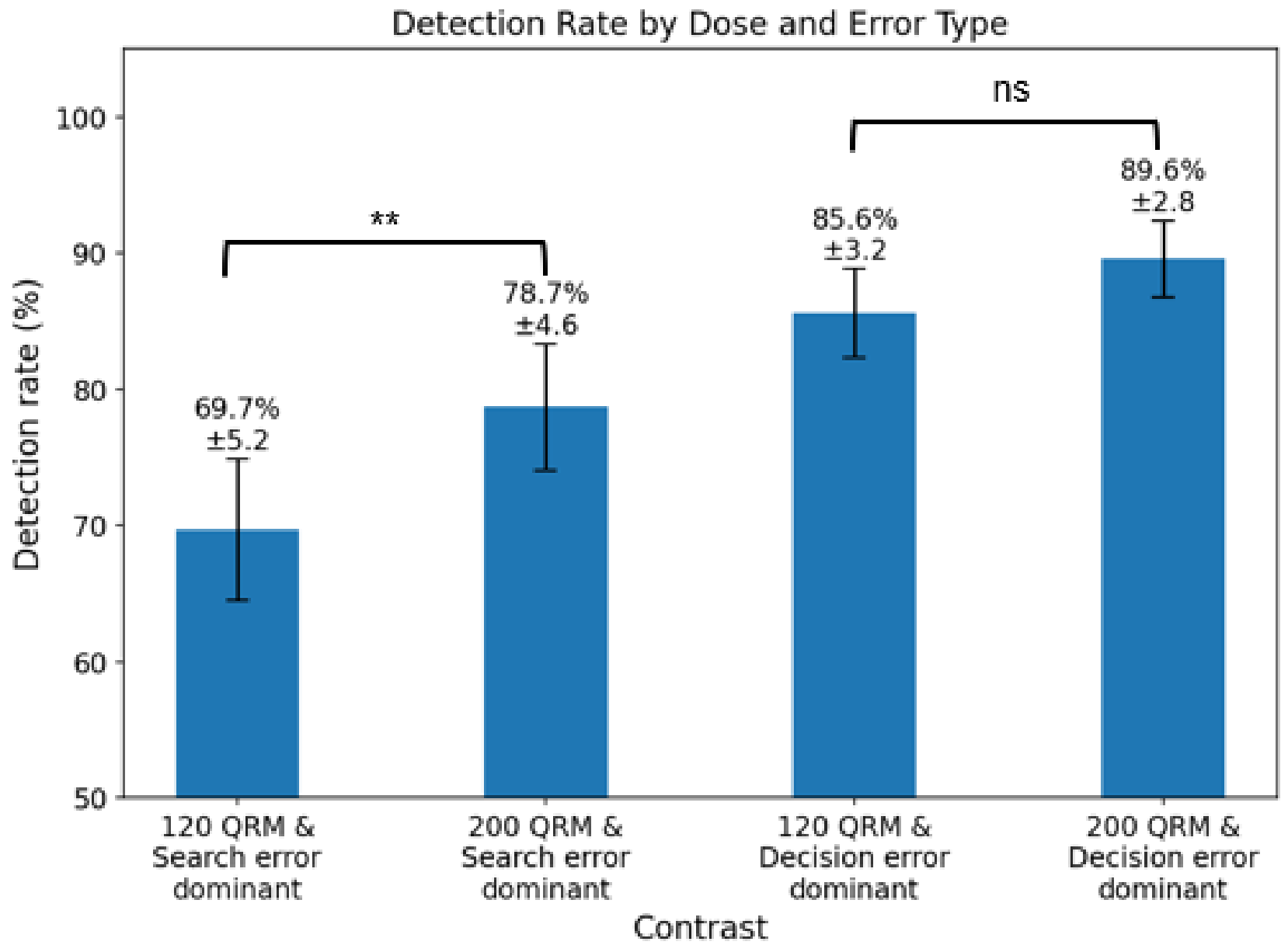

**Figure 6.** shows the detection rate for gaze threshold 1 second across dose levels and error categories: search-error-dominant and decision-error-dominant metastases, respectively, where ** represents p-value < 0.01 and ns represents not significant. QRM= quality reference milliampere-seconds.

**Table 4.** shows the chance of detection at 200 QRM vs 120 QRM, where CI is confidence interval. QRM= quality reference milliampere-seconds

| Lesion category | Gaze threshold (sec) | Odds ratio (CI) | p-value |
|---|---|---|---|
| Search-error-dominant | 2 | 1.62 (1.19, 2.21) | 0.002 |
| Decision-error-dominant | 2 | 1.34 (0.80, 2.24) | 0.25 |
| Search-error-dominant | 1 | 1.60 (1.17, 2.19) | 0.003 |
| Decision-error-dominant | 1 | 1.44 (0.94, 2.21) | 0.08 |

## Discussion

In this study, we investigated how dose reduction impacts lesion detection for hepatic metastases when they were categorized as search-error-dominant and decision-error-dominant metastases based on eye tracking gaze times. Our findings show that while overall detection decreases at lower dose, this decline is not uniform across error types. Specifically, search-error-dominant metastases showed greater sensitivity to dose reduction, whereas decision-error-dominant metastases were relatively less affected (22).

It has long been known that higher radiation dose significantly improves lesion detection because it reduces image noise. In our study, too, increasing dose from 120 to 200 QRM for search-error-dominant metastases at gaze threshold 2 seconds was associated with an increase in hepatic metastases detection (OR = 1.62, 95% CI: 1.19 - 2.21). The decision-error-dominant metastases form a minority of the errors (Table 1).

These findings align with the conceptual framework that lesion detection involves both visual search and decision-making processes. Reduced radiation dose primarily degrades image quality, making metastases harder to visually locate, thereby increasing search errors. In contrast, decision errors arise after the lesion has already been fixated upon and are therefore less dependent on image quality. This explains why decision-error-dominant metastases showed minimal change in detection frequency across dose levels.

From a clinical perspective, these results suggest that dose optimization strategies should consider not only overall detection performance but also the underlying mechanisms of error. If dose reduction disproportionately increases search errors, techniques that improve lesion conspicuity, such as advanced reconstruction algorithms or AI-assisted detection tools, may help mitigate this effect. On the other hand, decision errors may require different interventions, such as targeted training, improved reporting strategies, or decision support systems.

Decision errors are often ignored in simple models of lesion detectability, and they are essentially absent in phantom studies where readers know what lesion they are looking for, and the only task is to find it. However, decision errors occur in real-world settings, and their existence may help to explain the non-inferiority of small dose reductions for liver metastasis detection, or the underwhelming benefit of iterative reconstruction for metastasis detection (6). In most studies of this topic, a group of readers interprets images at multiple dose levels or reconstruction techniques, and these readers attempt to detect a population of metastases that are difficult to find. But within this population, there is a subpopulation of metastases that are decision-error-dominant, not search-error-dominant. Misses in this subpopulation occur essentially at random, without regards to manipulation of radiation dose or reconstruction technique. Decision-error-dominant metastases therefore reduce statistical power because they are unaffected by dose. Future studies, perhaps examining the ability of AI-powered reconstruction to maintain metastases detectability at reduced dose, should be designed with either a larger number of readers or metastases to compensate for the reduced power, or through selecting exams so that they preferentially contain search-error-dominant metastases rather than decision-error-dominant metastases.

One limitation of this study is that the dataset consists only of liver metastases, which may limit generalizability to other lesions or anatomical regions. Despite this limitation, this work provides a novel integration of eye tracking data with dose-dependent reader performance data, offering an improved understanding of why detection performance degrades at lower radiation dose. By linking gaze behavior with detection outcomes, we demonstrate that not all misses are equivalent and that distinguishing between search-error-dominant and decision-error-dominant metastases provides meaningful insight into radiologist performance.

## Conclusion

In this study, we demonstrated that missed liver metastases at lower CT radiation dose can be better explained by separating errors into search and decision errors using eye tracking data. We observed that search-error-dominant metastases, typically metastases with low contrast and limited conspicuity, are significantly affected by dose reduction, whereas decision-error-dominant metastases are missed at similar rates across dose levels. These findings highlight that not all missed metastases are equally sensitive to radiation dose reduction. Understanding the balance between perceptual and cognitive errors may help guide dose optimization and reader training strategies in CT interpretation.